\documentclass[onecolumn]{article}
\usepackage[utf8]{inputenc}
\usepackage{float}
\usepackage{hyperref}
\usepackage[frozencache]{minted}
\usepackage{graphicx}
\usepackage{verbatimbox}
\usepackage{array}
\usepackage{amsmath}
\usepackage{braket}
\usepackage{bm}
\usepackage[sorting=none,citestyle=numeric-comp]{biblatex}
\usepackage{appendix}
\usepackage{geometry}
\usepackage{authblk}

\usepackage{listings}
\usepackage{xcolor}
\usepackage{tcolorbox}

\definecolor{codegreen}{rgb}{0,0.6,0}
\definecolor{codegray}{rgb}{0.5,0.5,0.5}
\definecolor{codepurple}{rgb}{0.58,0,0.82}
\definecolor{codered}{rgb}{1.,0,0}
\definecolor{codepurple}{rgb}{0.58,0,0.82}
\definecolor{backcolour}{rgb}{0.95,0.95,0.92}
\definecolor{backgreen}{rgb}{0.90,0.96,0.94}

\setminted[python]{linenos=true,python3=true, bgcolor=backcolour, samepage=true}
\setmintedinline[python]{bgcolor={}, python3=true, breaklines=true}

\def \githubURL{https://github.com/goodchemistryco/Tangelo}
\def \notebooksURL{\githubURL/tree/main/examples}
\def \QEMISTCloudURL{https://goodchemistry.com/qemist-cloud}
\def \GCcontactURL{https://goodchemistry.com/contact}

\hypersetup{
    colorlinks = true,
    linkcolor = [rgb]{1.0, 0.5, 0.1}, 
    urlcolor = [rgb]{1.0, 0.5, 0.1},
    linkbordercolor = {white},
}

\lstdefinestyle{mystyle}{
    backgroundcolor=\color{backcolour},   
    keywordstyle=\color{codegreen},
    stringstyle=\color{codered},
    numberstyle=\tiny\color{codegray},
    basicstyle=\ttfamily\small,
    breakatwhitespace=false,         
    breaklines=true,                 
    captionpos=b,                    
    keepspaces=true,                 
    numbers=left,                    
    numbersep=5pt,                  
    showspaces=false,                
    showstringspaces=false,
    showtabs=false,                  
    tabsize=2
}

\title{Tangelo: An Open-source Python Package for End-to-end Chemistry Workflows on Quantum Computers}
\author[1,\footnote{Corresponding author: valentin@goodchemistry.com}]{Valentin Senicourt}
\author[1]{James Brown}
\author[1]{Alexandre Fleury}
\author[3]{Ryan Day}
\author[1]{Erika Lloyd}
\author[2]{Marc P. Coons}
\author[1]{Krzysztof Bieniasz}
\author[1]{Lee Huntington}
\author[2]{Alejandro J. Garza}
\author[3]{Shunji Matsuura}
\author[1]{Rudi Plesch}
\author[1]{Takeshi Yamazaki}
\author[1]{Arman Zaribafiyan}
\affil[1]{\textit{Good Chemistry Company, 200-1285 West Pender Street Vancouver, BC, V6E 4B1, Canada}}
\affil[2]{\textit{Dow Inc., Core R\&D, Chemical Science, 1776 Building, Midland, MI, 48674, USA}}
\affil[3]{\textit{1QB Information Technologies (1QBit), 200-1285 West Pender Street Vancouver, BC, V6E 4B1, Canada }}

\begin{document}

\maketitle

\begin{abstract}
    Tangelo (\href{https://github.com/goodchemistryco/Tangelo}{https link}) is an open-source Python software package for the development of end-to-end chemistry workflows on quantum computers, released under Apache 2.0 license. It aims to support the design of successful experiments on quantum hardware, and to facilitate advances in quantum algorithm development. The software enables quick exploration of different approaches by assembling reusable building blocks and algorithms, with the flexibility to let users introduce their own. Tangelo is backend-agnostic and enables switching between various backends (Braket, Qiskit, Qulacs, Azure Quantum, QDK, Cirq...) with minimal changes in the code. The package can be used to explore quantum computing applications such as open-shell systems, excited states, or more industrially-relevant systems by leveraging problem decomposition at scale. This paper outlines the design choices, philosophy, and main features of Tangelo.
\end{abstract}

\section{\label{sec:intro} Introduction}

It is an exciting time to be working in the field of quantum computing, as we are witnessing rapid advancements in both software and hardware developments. The community still has to address many challenges before this technology can be employed to reliably solve industrially-relevant problems. A potential area of promise is the study of chemical systems and materials science, which is the main focus of Tangelo.

Indeed, many of the methods that will give quantum devices advantage over their classical counterparts are yet to be identified. For this purpose, we need tools that facilitate the exploration and comparison of different approaches, as well as the design of successful experiments on quantum devices. We need to design practical end-to-end workflows able to return an answer within acceptable accuracy while meeting the computational resource constraints, such as number of qubits, quantum gates, or measurements.

Currently, studying basic molecular systems on quantum computers is impossible when naively embedding the whole problem on quantum hardware and simply executing a straightforward quantum algorithm.
Leveraging pre- and post-processing techniques as well as insights from classical calculations remains necessary in order to make a problem  computationally tractable while maintaining accuracy.  Assembling the different building-blocks to form and explore workflows that meet these constraints is where Tangelo strives to be of help. This document covers the features, API, and philosophy of this software package.\\

Section 2 of this document presents an overview of what we mean by end-to-end workflows, and looks at what challenges the different building blocks present in this package attempt to tackle. Section 3 shows how Tangelo's API and data-structures work in practice, covering the different steps of end-to-end workflows relating to hardware experiments. Section 4 details our philosophy, as well as the various processes behind the open-sourcing, distribution, testing and development of Tangelo. Finally, section 5 elaborates on our goal to develop a community and engage in collaborative work, providing examples of topics that could enrich these tools and an overview of our current roadmap.

\section{Overview of the End-to-end Workflow\label{sec:overview}}

\begin{figure}
    \centering
    \includegraphics{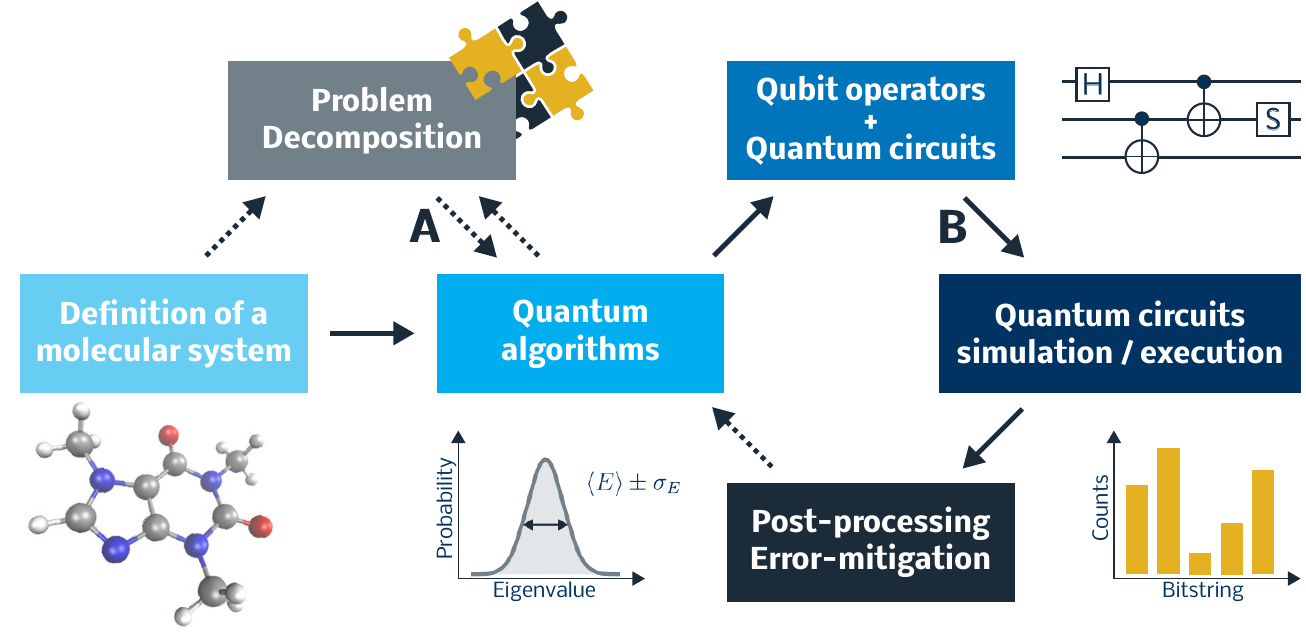}
    \caption{Simplified representation of a quantum computing workflow for quantum chemistry problems, highlighting the main steps of the process. Dotted arrows represent optional steps. \textbf{A} Problem decomposition methods can be paired with quantum algorithms, in order to lower computational resource requirements of a use case. \textbf{B} Pre-processing tools aiming at designing efficient measurement protocols, or maximizing accuracy of results (circuit compilation, optimizations, etc).}
    \label{fig:quantum_workflow}
\end{figure}

Tangelo facilitates the design of quantum computing experiments, with chemical systems as the main target. An overview of how end-to-end workflows are used for this purpose is illustrated in Figure~\ref{fig:quantum_workflow}. These workflow encompass all steps in the simulation, from inputting the molecular data to returning post-processed experimental results. They can be complex, and each step involves its own interesting opportunities for innovation, which can have strong impact on resource requirements.

\subsection{Defining a Molecular System}

Molecular systems can be defined in Tangelo by providing the cartesian coordinates for the different atoms comprising the system, as well as additional information such as the spin, charge or basis set considered. Cartesian coordinates come in different formats (XYZ, MOL2, PDB, ...) containing specialized features relevant to the simulation method used (respectively quantum, atomistic or biochemical modelling). The XYZ format is the one used in Tangelo inputs, although easy conversion from other molecular formats is possible through projects such as the Open Babel initiative~\cite{openbabel}.

Most quantum algorithms take as input a Hamiltonian.  The molecular electronic Hamiltonian can be written in second-quantization as
\begin{equation}\label{eq.ham}
    H = \sum_{pq}h_{pq} a_p^{\dagger}a_q + \frac{1}{2}\sum_{pqrs}h_{pqrs}a_p^{\dagger}a_r^{\dagger}a_s a_q,
\end{equation}
where $a_p$ is the fermionic lowering operator for orbital $p$ and $h_{pq}, h_{pqrs}$ are the 1- and 2-electron integrals in chemist's notation. One typically works in a finite basis of atomic orbitals and solves the mean-field Hartree-Fock problem to determine a suitable set of molecular orbitals (which can be expressed as a linear combination of the atomic orbitals). The 1- and 2-electron integrals can then be transformed to this molecular orbital basis, which then provides a suitable representation for the molecular electronic Hamiltonian in that basis. The Hartree-Fock calculations and integral transformation steps are currently handled by third-party packages such as PySCF~\cite{pyscf1,pyscf2}. Tangelo's modular design allows us to support alternative packages in the future.

\subsection{Algorithms}

This package contains implementations of both quantum and classical algorithms for quantum chemistry, built upon the various lower-level functionalities designed in the toolboxes. They are further classified depending on their nature, or the feature of a molecular system that they aim to compute. The goal of this module is to provide a collection of algorithms that users can reuse, in order to quickly put together end-to-end chemistry workflows, and combine and compare different approaches.

Quantum algorithms attempt to solve a problem by recasting it into quantum circuits and qubit operators, and running these circuits on a backend such as a simulator or a quantum device. The interface in Tangelo enables easy estimation of computational resources (number of qubits, gates, etc.), ease of use by separating the initialization of underlying data-structures from the execution of the algorithm, and also gives users the ability to pass customizing options for low-level control. So far, Tangelo features various flavours of the Variational Quantum Eigensolver (VQE) ~\cite{VQE,qubitwise2}, such as ADAPT-VQE for iterative circuit building~\cite{ADAPT-VQE}, and State-Averaging VQE with orbital optimization for excited states~\cite{SAOOVQE}. It also features the Quantum Imaginary Time-Evolution (QITE)~\cite{QITE} algorithm to obtain an eigenstate of a Hamiltonian. Other algorithms include (controlled) time-evolution and Quantum Fourier Transform (QFT), which facilitate computing the energy of a state with phase estimation~\cite{Time-Evolution} or projecting a state onto a certain symmetry~\cite{Sym-Prep} or energy~\cite{Rodeo}. Work on other quantum algorithms is underway.

Classical algorithms are provided for convenience, in order to acquire reference results and reflect on the performance of quantum algorithms when applied to concrete use cases. As the capabilities of quantum devices and simulators are currently limited, these classical algorithms tend to be able to deliver faster and more accurate results, and scale to larger systems. The list so far includes Full-Configuration Interaction (FCI), Coupled-Cluster Singles and Doubles (CCSD) ~\cite{CCSD,RevModPhys.79.291} and the semi-empirical solver MINDO3 ~\cite{MINDO3_1,MINDO3_2}.

The parameters for these algorithms can have significant impact on their accuracy, or their compute requirements. For example, one could study how freezing some electronic orbitals impacts the accuracy of the solution using the classical solvers available, before incorporating these insights into their quantum workflow. In general, we designed our algorithms in a modular and flexible way that provides users with a fine level of control, and the ability to integrate their own custom code. We hope to grow this collection of algorithms, in order to provide users with the tools to investigate and tailor various approaches to their use cases, and support them in running successful experiments on quantum devices.

\subsection{Problem Decomposition}

The main objective of problem decomposition (PD) techniques, also referred to as fragmentation techniques, is to lower computational resource requirements by decomposing the initial problem instances into a collection of subproblems. These are solved using fewer resources, and their results recombined to approximate the full solution. Examples of such algorithms include Density-Matrix Embedding Theory (DMET) ~\cite{DMETa, DMETb} and the Method of Increments (MI) ~\cite{MI1,MI2,MI3,Stoll:1992aa,Modll:1997aa,Stoll:2005aa,Friedrich:2007aa,Friedrich:2013a,Friedrich:2013b,Richard:2012,Zimmerman:2017ab,Zimmerman:2017ab,Zimmerman:2017ac,Zimmerman:2019aa,Eriksen:2018,Eriksen:2019a,Eriksen:2019b,Verma2021}. The subproblems can then be solved with a combination of the supported classical or quantum algorithms described in the previous section.

Naively embedding a whole problem and throwing it at a quantum computer is not practical, due to the limited capabilities of current quantum devices. Problem decomposition can allow users to go beyond toy problems and explore the impact of quantum computing on more industrially-relevant instances. It can combine quantum and classical algorithms, sometimes at different levels of theory, in order to solve the different subproblems generated, allowing users to introduce and study quantum computation as part of more complex use cases.

Problem decomposition is a common technique in chemical modelling, and it covers a broad swath of methods. The common theme in these methods is to partition the overall system into a system of interest, where an accurate description of the interactions must be used, and subsystems where cruder approximations between particles is sufficient. Some PD techniques are based on partitioning around electronic orbitals, taking into account electronic correlation between subsystems (e.g. the method of increments). Other partition systems based on atoms and bonds, creating molecular fragments, such as the well-known Quantum-Mechanical/Molecular-Mechanical methods ~\cite{warshel1976theoretical,mulholland2000ab, senn2009qm,groenhof2013introduction}. Finally, some PD techniques based on molecular fragments only treat the fragment in the system of interest explicitly, with the low-level subsystems approximated as a continuum solvent, such as the Polarizable Continuum Model (PCM)~\cite{pcm1,pcm2,pcm3}

\begin{figure}
    \centering
    \includegraphics{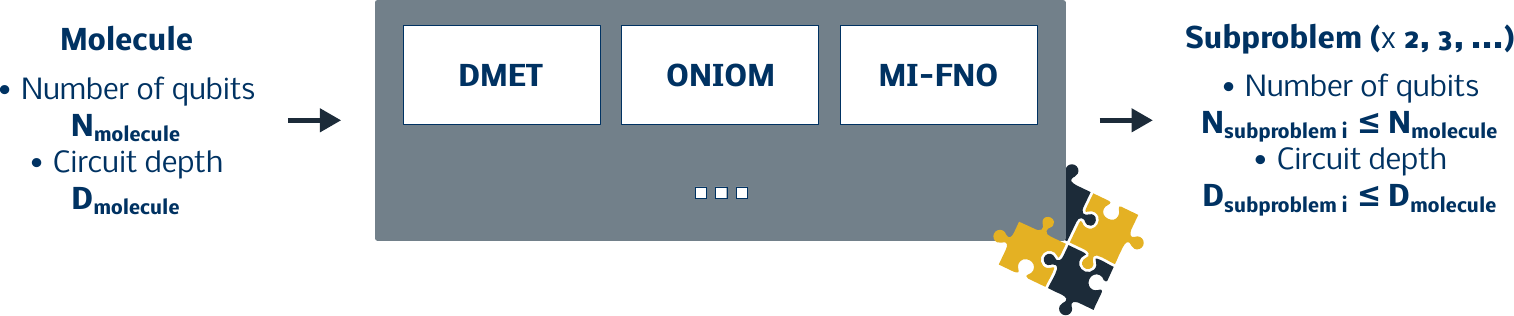}
    \caption{Problem decomposition aims at reducing computational resource requirements while maintaining the accuracy of the results. Paired with quantum algorithms, it may in particular make metrics such as number of qubits or quantum gates more tractable on existing hardware.}
    \label{fig:pd}
\end{figure}

The impact of problem decomposition on both computational resource requirements and accuracy depends on the use case, and is a topic of ongoing research~\cite{Verma2021,Yamazaki2018}. We describe in the following subsections a few of the PD methods currently supported in Tangelo.

\subsubsection{Density-Matrix Embedding Theory (DMET)}

DMET ~\cite{DMETa,DMETb} is a quantum embedding technique that breaks the molecule into fragments, defined by atoms and the corresponding localized basis functions. Each fragment is solved as an open quantum system with the fragment orbitals connected to the bath orbitals such that a new Hamiltonian is defined for each fragment. The Hamiltonian can be rewritten as
\begin{equation}
    H = \sum_{pq}\left[h_{pq}+\sum_{rs}\left[h_{pqrs}-h_{psrq}\right]D_{rs}^{\textrm{env},A}\right] a_p^{\dagger}a_q -\mu \sum_{p}a_p^{\dagger}a_p  +\sum_{pqrs}h_{pqrs}a_p^{\dagger}a_q^{\dagger}a_s a_r,
\end{equation}
where $D^{\textrm{env},A}$ is the density matrix for the environment of fragment $A$, $\mu$ is the chemical potential updated after each DMET cycle by $a(\sum_{A}N^{A}-N^{\textrm{tot}})$ where $a$ is a positive number, $N^{A}$ and $N^{\textrm{tot}}$ are the number of electrons in fragment $A$ and the whole system respectively. 

The DMET cycle can be described by the pseudocode shown in Figure~\ref{fig:DMET}. Solving each fragment Hamiltonian can be achieved with a quantum computer (simulated or hardware) or with a classical solver. 

\begin{figure}[!ht]
\begin{minted}[escapeinside=||,mathescape=true]{text}
COMPUTE the mean field
DECOMPOSE molecule into fragments
|\bf{while $\sum_{A} N^{A} \neq N^{tot}$}|
    DETERMINE the chemical potential
    |{\bf{for}} $A \in$ fragments|
        CONSTRUCT bath orbitals |$L_B$| for fragment |$A$|
        CONSTRUCT the fragment Hamiltonian |$H$|
        SOLVE Hamiltonian
        COMPUTE 1- and 2-RDMs and |$N^A$|
    COMPUTE the total energy
\end{minted}
\caption{The pseudocode for the DMET algorithm.\label{fig:DMET}}
\end{figure}

As DMET requires 2-RDMs for each fragment, a post Hartree-Fock method must be used for each fragment. The user is in control of the different atoms defining each fragment, and other parameters defining the behaviour of the algorithm.

\subsubsection{ONIOM}

Another available problem decomposition technique is ONIOM \cite{ONIOM_1,ONIOM_2,ONIOM_3,ONIOM}, a hybrid approach breaking down a molecular system into subproblems that can be solved with different levels of theory. In particular, this class of algorithms enables users to tackle larger systems, targeting the most important parts of the system with expensive but accurate algorithms using a high-level theory, while treating the rest of the system with a lower-level theory.

Eq.~\eqref{eq.ham} is broken into fragments, and disregards the interaction terms between the fragments. The total energy $E_{\text{ONIOM}}$ is computed as
\begin{equation}
    E_{\text{ONIOM}} = E_{\text{All}}^{\text{Low}} + \sum_{i=1}^N (E_{\text{Fragment}_i}^{\text{High}} - E_{\text{Fragment}_i}^{\text{Low}}),
\end{equation}
where $E_{\text{All}}^{\text{Low}}$ is the energy of the whole system at a low level of theory, and $E_{\text{Fragment}_i}^{\text{High}}$ and $E_{\text{Fragment}_i}^{\text{Low}}$ is the energy of fragment $i$ at a high- and low-level of theory respectively. High- and low-level of theory are referring here to the method's accuracy towards computation of total molecular energy. A 2-RDM is not required to calculate the total energy so Hartree-Fock or semi-empirical methods can be used for the low-level calculations.

ONIOM use cases range from solution chemistry, as illustrated in the next section, to nanomaterials simulation, biochemistry applications and beyond~\cite{ONIOM}. Computational scientists have taken this approach to study enzymatic reaction mechanisms, visualize protein folding or tackle other problems that remain out of reach for pure quantum chemistry methods.

\subsubsection{Method of Increment with Frozen Natual Orbitals (MI-FNO)}

The method of increments~\cite{MI1,MI2,MI3,Stoll:1992aa,Modll:1997aa,Stoll:2005aa,Friedrich:2007aa,Friedrich:2013a,Friedrich:2013b,Richard:2012,Zimmerman:2017ab,Zimmerman:2017ac,Zimmerman:2019aa,Eriksen:2018,Eriksen:2019a,Eriksen:2019b,Verma2021} (MI) can be used to express the correlation energy of a molecular system as a truncated many-body expansion in terms of orbitals, atoms, molecules, or fragments. In Tangelo, the correlation energy of the system is expanded in terms of occupied orbitals, and MI is employed to systematically reduce the occupied orbital space. Namely, the correlation energy can be expressed in terms of $n$-body increments ($\epsilon_{i} $, $\epsilon_{ij}$, $\epsilon_{ijk}$, $\epsilon_{ijkl}$), and so on, as 
\begin{align}\label{Ec_dec}
	E_\text{c} &= E_{\text{exact}} - E_{\text{HF}}\nonumber\\
	&=\sum_i \epsilon_{\text{i}} + \sum_{i>j} \epsilon_{ij} + \sum_{i>j>k} \epsilon_{ijk} +\sum_{i>j>k>l} \epsilon_{ijkl} + \ldots
\end{align}
Here, $\epsilon_{i} $, $\epsilon_{ij}$, $\epsilon_{ijk}$, and $\epsilon_{ijkl}$ are, respectively, the one-, two-, three-, and four-body increments which are defined as
\begin{align}
	\epsilon_{i} &= E_\text{c}(i),\\
	\epsilon_{ij} &= E_\text{c}(ij) - \epsilon_{i} - \epsilon_{j},\\
	\epsilon_{ijk} &= E_\text{c}(ijk) - \epsilon_{ij} - \epsilon_{ik} - \epsilon_{jk} - \epsilon_{i} - \epsilon_{j} - \epsilon_{k},\\
	\epsilon_{ijkl} &= E_\text{c}(ijkl) - \epsilon_{ijk} - \epsilon_{ijl} - \epsilon_{jkl} - \cdots,\\
	&\vdots\nonumber
\end{align}
in which $E_\text{c}(i)$ denotes the correlation energy of the increment $i$, $E_\text{c}(ij)$ denotes the correlation energy of the increment $i,j$, and so on.

At the same time, the virtual orbital space is reduced by using the so-called frozen natural orbital (FNO) approach~\cite{FNO_Davidson,FNO_bartlett,FNO_taube1,FNO_taube_lamdaCCSD}. Namely, in this scheme, the virtual space is spanned by natural orbitals of the one-particle density matrix from second-order, many-body perturbation theory and the virtual orbitals whose occupation numbers (i.e. eigenvalues of the density) fall below a threshold are discarded. In this way, a method referred to as the MI-FNO approach is available for the systematic reduction of both the occupied space and the virtual space in quantum chemistry simulations.

We refer the reader to our \href{\notebooksURL/mifno.ipynb}{tutorial notebook} for a gentle introduction to MI-FNO and how it can be leveraged in Tangelo, and to~\cite{Verma2021} for a more comprehensive description of the approach.

\subsection{Quantum Circuit Simulation and Execution}

\subsubsection{Backend-agnosticism}

The idea of being backend-agnostic is that users only have to write their algorithm once, and can then easily run it on various simulators and quantum devices with minimal changes. All quantum algorithms available in Tangelo are written using generic data-structures that are backend-agnostic: they are thus not tied to a specific platform. Assembling gates and forming complex circuits in Tangelo is very straightforward and pythonic, as we demonstrate in section 3. 

The cornerstone of our approach relies on format conversion functions, which are able to convert the generic ``abstract'' format Tangelo uses to represent a quantum circuit into an object or simple string that can be used by the target backends. Users are able to use these functions directly to export a quantum circuit in their desired format (including QASM), or provide it as input to the API of the framework of their choice, such as local code or quantum cloud services. This feature is particularly important, as it ensures compatibility with the growing number of projects that study questions related to quantum circuits, such as their simulation, or compilation/optimization for specific architectures.

Tangelo provides a unified interface to prominent quantum circuit simulation packages developed by the community, such as Cirq~\cite{Cirq}, Qiskit~\cite{Qiskit}, Qulacs~\cite{Qulacs}, Braket~\cite{Amazon_Braket} and QDK~\cite{qdk}. By simply changing one word in their code, users can switch between different simulators. Likewise, other parameters enable users to run their custom noisy simulations, or decide how many shots (measurements) should be used to compute the output of the simulation. This layer of abstraction effectively decouples quantum chemistry from quantum circuit simulation, and allows users to focus on the former, by not requiring them to learn and write code specific to a particular platform.

We look forward to integrating new promising backends and platforms displaying outstanding performance or unique features.

\subsubsection{Connection to Quantum Devices}

Tangelo can reach a variety of quantum devices, thanks to the format conversion functions it supports. This allows users to access quantum cloud providers such as Braket or Azure Quantum, to directly use their favorite hardware provider's API, or to use the convenience functions in \href{\QEMISTCloudURL}{QEMIST Cloud}, Good Chemistry's platform enabling high-accuracy quantum chemistry simulations in the cloud.

It is possible for users to extract objects such as quantum circuits or qubit operators at various steps of a quantum algorithms, and use these as inputs for an experiment on a quantum device. Currently, the quantum algorithms written in Tangelo only streamline simulators as backends. We are however open to providing support for running whole algorithms on quantum devices in the future, as soon as the community deems it interesting.

\subsection{Pre- and Post-processing for Quantum Experiments}




The output of a quantum computer is a histogram of measurements corresponding to the different outcomes observed, usually expressed as bitstrings. The cost and duration of a quantum experiment is roughly linear with the number of shots used to build such histograms, which also correlates with the accuracy of the results.

Pre-processing encompasses several steps preceding execution of the circuits on a quantum device, such as defining efficient measurement protocols or compiling a quantum circuit for a particular architecture, among others. Tangelo provides a collection of features to help users design efficient measurement protocols, with the goal of lowering the cost of an experiment while preserving accuracy as much as possible. For example, such features encompass grouping strategies, which identifies the smallest set of measurement bases in which circuits need to be run in order to derive all the values required to compute the end results, while extracting the most information from each measurement.

As devices only support a set of native quantum gates, it is also important that quantum circuits are compiled and optimized for the target architecture. The insights provided during this step may help further reduce the amount of measurements needed for the experiment, or increase the accuracy of the end result. Due to its backend-agnostic approach, Tangelo is compatible with third-party tools developed by the quantum community and hardware providers, in order to tackle these challenges. We look forward to supporting better integration with these frameworks, as well as providing features directly in Tangelo when relevant.

Post-processing takes as input histograms returned by the quantum device and can derive quantities of interest, as well as their error bars to account for statistical effects. This process may include noise-mitigation techniques, which attempt to improve the derived values either by using insights of the molecular system, or characterized properties of the quantum hardware. Tangelo provides general and chemistry-inspired post-processing techniques, to assist users with obtaining successful experiment results, and aims to provide more in the near future.



\section{\label{sec:data_structures}API Overview Through End-to-end Examples}

This section demonstrates how Tangelo's API works, through some examples used in previous works, including an end-to-end pipeline featured in reference~\cite{DMET_H10}. We elaborate on the implementation and interface of the building blocks used for each step. These buildings blocks can be low-level functions drawn from one of the many toolboxes in the package, or more complex ones built on top of them, such as a whole quantum algorithm, combined with a problem decomposition technique.

Tangelo currently contains high-level submodules listed in the table below, matching the folder hierarchy of the package.\\

\newcolumntype{L}{>{\arraybackslash}m{13cm}}
\begin{tabular}{|l|L|}
  \hline
  \texttt{algorithms} & Quantum and classical algorithms for quantum chemistry, further subdivided depending on their nature and goal. Built upon the various functionalities designed in the toolbox modules.\\ 
  \hline
  \texttt{problem\_decomposition} & Methods for decomposing a molecular system into a collection of smaller subproblems and reconstructing its properties by solving them. Designed to reduce resource requirements of the input problem, to adhere to computational constraints. Can be paired with the algorithms defined in the algorithms module.\\ 
  \hline
  \texttt{toolboxes} & Low-level building blocks supporting the development of algorithms and end-to-end workflows, covering topics such as frozen orbitals, qubit mappings, ansatz circuits, error-mitigation, qubit terms grouping, etc.\\ 
  \hline
  \texttt{linq} & Interface to backend-agnostic representation and operations on quantum circuits, format conversion and wrappers to facilitate connection to diverse simulators and quantum devices.\\ 
  \hline
  \texttt{helpers} & Miscellaneous helper functions.\\ 
  \hline
\end{tabular}
\\

\subsection{Inputs}

A molecular system can be defined by passing its atoms' cartesian coordinates (xyz), charge (q), spin, a basis set and the list of frozen molecular orbitals.

In order to tackle this system with methods working in second quantization, such as the popular Variational Quantum Eigensolver (VQE), we use this information to instantiate an object of the \mintinline{python}{SecondQuantizedMolecule} class. Please note that Tangelo is not restricted to algorithms and data-structures operating in second quantization: the API will be naturally extended as more approaches are introduced to the codebase, such as first-quantization.

\begin{minted}{python}
from tangelo import SecondQuantizedMolecule

xyz = [['H', (0.0, 1.780, 0.0)], ['H', (-1.046, 1.44, 0.0)], 
       ['H', (-1.693, 0.55, 0.0)], ['H', (-1.693, -0.55, 0.0)], 
       ['H', (-1.046, -1.44, 0.0)], ['H', (0.0, -1.78, 0.0)], 
       ['H', (1.046, -1.44, 0.0)], ['H', (1.693, -0.55, 0.0)], 
       ['H', (1.693, 0.55, 0.0)], ['H', (1.046, 1.44, 0.0)]]

mol = SecondQuantizedMolecule(xyz, q=0, spin=0, basis="minao", frozen_orbitals=None)
\end{minted}

This instantiation of \mintinline{python}{SecondQuantizedMolecule} includes the computation of the mean-field solution. Currently, these conventional quantum chemistry calculations rely on the PySCF~\cite{pyscf1,pyscf2} python package; we may support other computational chemistry packages in the future. Restricted closed- and open-shell Hartree-Fock are currently supported automatically by changing the \mintinline{python}{spin} argument. This is the starting point for post-HF calculations for introducing electronic correlation. Besides general molecular information, \mintinline{python}{SecondQuantizedMolecule} contains data about (in)active molecular and spin orbitals.

\subsection{Quantum Chemistry Algorithms}

\subsubsection{\label{subsec:classical_algorithms}Classical Algorithms}

Classical algorithms can be useful to obtain reference results, or to get quick insights about the system (consequences of freezing orbitals, etc) that can be later leveraged during the exploration of quantum approaches. They are here for convenience and support, and allow users to also combine classical and quantum solvers in more elaborate approaches. For instance, \mintinline{python}{SecondQuantizedMolecule} can be passed to instantiate classical solvers such as \mintinline{python}{FCISolver} and \mintinline{python}{CCSDSolver}. The  \mintinline{python}{simulate} method can then be used to compute the ground state energy.

\begin{minted}{python}
from tangelo.algorithms import FCISolver, CCSDSolver

fci_energy = FCISolver(mol).simulate()
ccsd_energy = CCSDSolver(mol).simulate()
\end{minted}

These classical solvers also allow you to obtain 1- and 2-RDMs easily with the \mintinline{python}{get_rdm} method. Both \mintinline{python}{CCSDSolver} and \mintinline{python}{FCISolver} freeze orbitals as directed by the user when obtaining classical solutions. 
The implementation of those solvers currently leverages PySCF~\cite{pyscf1,pyscf2}.

\subsubsection{Quantum Algorithms}

Tangelo provides implementations of different quantum algorithms, including variational approaches such as VQE~\cite{VQE} and ADAPT-VQE~\cite{ADAPT-VQE}. Tangelo is neither restricted to variational approaches nor second-quantization, and we look forward to supporting a broader collection of methods. 
The API of quantum algorithms is similar to the one for classical algorithms, with extra methods:

\begin{itemize}
\item \mintinline{python}{build} is a method that initializes all internal objects in the solver, according to the options passed by the user. 

\item
\mintinline{python}{get_resources} exposes a summary of computational resource requirements, such as metrics characterizing the current underlying quantum objects (circuit, Hamiltonians...) and other quantities of interest (for VQE, the number of variational parameters in the ansatz circuit gives us a sense of how difficult the classical optimization will be). It can be called right after \mintinline{python}{build} to let users access and assess the underlying objects and metrics (such as computational resource requirements) for their quantum algorithm without having to simulate it, or after simulation as well.

\item
\mintinline{python}{simulate} can be called after \mintinline{python}{build} has completed, and will run the quantum algorithm, simulating quantum circuits on the backend of your choice. Currently, we do not streamline the process to run entire algorithms on actual quantum devices.
\end{itemize}

The method \mintinline{python}{get_resources} can be called at any point after \mintinline{python}{build} has been called, and will return information based on the current state of the solver object. For instance, if called right after \mintinline{python}{build}, it will report values related to the initial objects used by the algorithm, before any quantum circuit has been simulated. If called after \mintinline{python}{simulate} has been run, it will return something different, such as the final ansatz circuit and qubit operators used in VQE, after classical optimisation is complete. Depending on the algorithm, more methods may be available. Such methods may allow users to compute intermediary quantities, instead of simulating the whole algorithm. For instance, VQE allows the direct computation of energy or RDMs, according to input variational parameters provided by the user (i.e., a single energy estimation). Please refer to the API and the \href{\notebooksURL}{example notebooks} to learn about some of these useful features.

\begin{figure}
    \centering
    \includegraphics{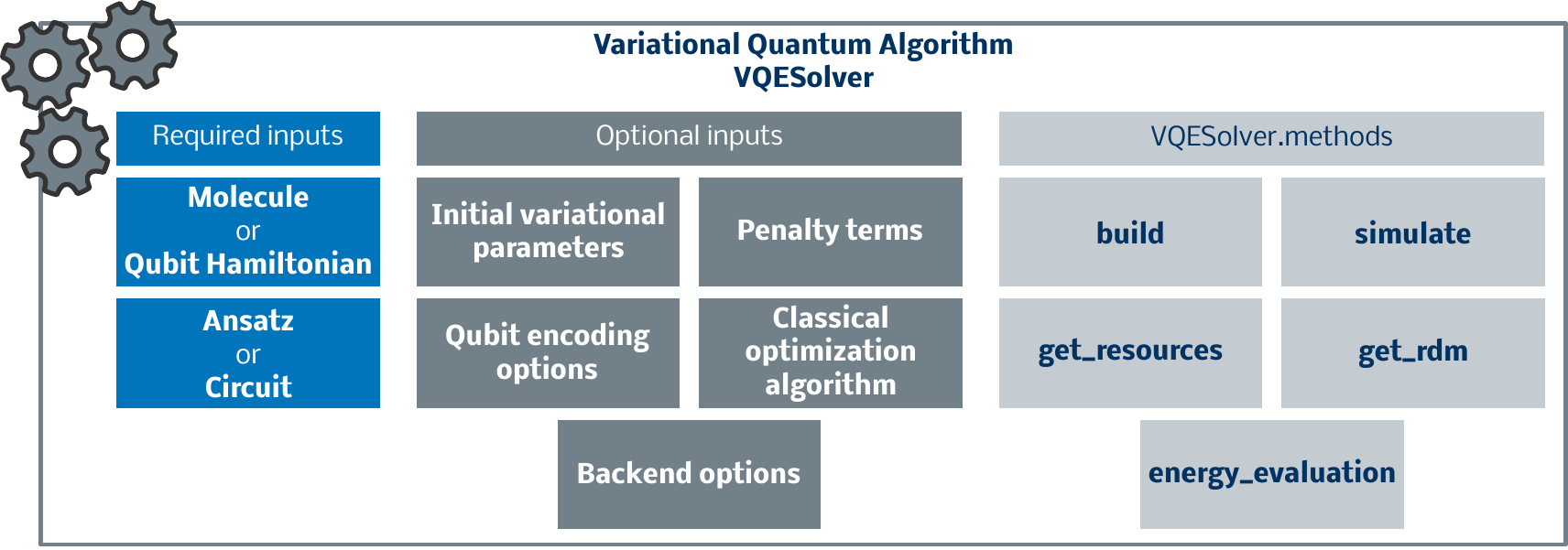}
    \caption{Main options for Tangelo \mintinline{python}{VQESolver}. VQE aims to approximate an eigenvalue by classically minimizing an expectation value, with regards to variational parameters used in an arbitrary quantum circuit (the ``ansatz''). Users have access to various options to customize the behaviour of the algorithm, and can both compute and retrieve values of interest by using the attributes and methods of the \mintinline{python}{VQESolver} object. For more details, check out our \href{\notebooksURL}{VQE tutorial notebooks}.}
    \label{fig:vqesolver}
\end{figure}

We strive to develop a very flexible approach that can both be used ``as-is'', but also provide users with the ability to control the behaviour of algorithms through options dictionaries. These options can significantly alter the behaviour and computational cost of a given algorithm, provide users with fine control over it, or even facilitate drop-in of their own custom code. For VQE, some notable parameters include \mintinline{python}{"qubit_mapping"} (e.g  Jordan-Wigner, Bravyi-Kitaev, symmetry-conserving-Bravyi-Kitaev, Jiang-Kalev-Mruczkiewicz-Neven~\cite{JKMN},...), initial variational parameters, or the choice of \mintinline{python}{"ansatz"}. Tangelo provides \mintinline{python}{BuiltInAnsatze} such as UCCSD~\cite{UCCSD1,UCCSD2,UCCSD3}, k-UpCCGSD~\cite{k-UpCCGSD}, HEA~\cite{HEA}, QCC~\cite{QCC}, RUCC~\cite{RUCC} and Variationally Scheduled Quantum Simulation (VSQS)~\cite{VSQS}, but users can define and pass their own ansatz objects, or even simply pass any variational quantum circuit defined in Tangelo.
We highlight some of these functionalities for the QCC ansatz with an illustrative example in the following paragraphs, after briefly reviewing the methodology. The details for utilizing the HEA ansatz and the \mintinline{python}{Ansatz} class, the latter of which enables the user to define custom ansatz, are provided in a \href{\notebooksURL/vqe_custom_ansatz_hamiltonian.ipynb}{dedicated tutorial notebook}. More generally, each of the ansatz available in the \mintinline{python}{ansatz_generator} toolbox is documented.

The QCC method is a variational approach that augments a mean-field wave function with electron correlation effects.
The QCC energy is obtained by optimizing an energy functional,
\begin{equation}\label{eq.qcc1}
  E_{\text{QCC}}(\bm{\tau},\bm{\Gamma}) = \bra{\bm{\Psi}(\bm{\tau},\bm{\Gamma})} \hat{H} \ket{\mathbf{\Psi}(\bm{\tau},\bm{\Gamma})},
\end{equation}
for two sets of variational parameters $\bm{\tau}$ and $\bm{\Gamma}$, that describe the correlated and mean-field wave functions, respectively.
In Eq.~\ref{eq.qcc1}, $\ket{\mathbf{\Psi}(\bm{\tau},\bm{\Gamma})}$ is the variational QCC wave function that takes form as
\begin{equation}\label{eq.qcc2}
 \ket{\mathbf{\Psi}(\bm{\tau},\bm{\Gamma})} = \hat{U}(\bm{\tau}) \ket{\bm{\Omega}(\bm{\Gamma})},
\end{equation}
and $\hat{H}$ is a molecular Hamiltonian that has been mapped to qubits (i.e., a qubit Hamiltonian).
For the QCC method, electron correlation is derived from a mean-field wave function $\ket{\bm{\Omega}(\bm{\Gamma})}$ through application of a unitary operator $\hat{U}(\bm{\tau})$.
The QCC unitary operator is constructed from a set of $n_{g}$ $\bm{\tau}$ parameters together with a set of Pauli word generators ${\hat{P}_{k}}$, the form of which is
\begin{equation}\label{eq.qcc3}
  \hat{U}(\bm{\tau}) = \prod_{k}^{n_{g}}\ \exp\left(-\frac{\mathrm{i} \tau_{k} \hat{P}_{k}}{2}\right).
\end{equation}
The mean-field wave function $\ket{\bm{\Omega}(\bm{\Gamma})}$ is expressed as a tensor product of $n_{q}$ single-qubit states $\ket{\bm{\omega}(\theta_{j},\phi_{j})}$ that are each specified by a pair of Bloch sphere polar and azimuthal angles $\theta_{j}$ and $\phi_{j}$:
\begin{equation}\label{eq.qcc4}
  \ket{\bm{\Omega}(\bm{\Gamma})} = \bigotimes_{j}^{n_{q}} \ket{\bm{\omega}(\theta_{j},\phi_{j})}.
\end{equation}
Each single-qubit state $\ket{\bm{\omega}(\theta_{j},\phi_{j})}$ is represented as a linear combination of the canonical Bloch sphere basis vectors $\ket{\bm{0}}$ and $\ket{\bm{1}}$:
\begin{equation}\label{eq.qcc5}
  \ket{\bm{\omega}(\theta_{j},\phi_{j})} = \cos\left(\frac{\theta_{j}}{2}\right) \ket{\bm{0}_{j}} + \text{e}^{\mathrm{i} \phi_{j}}\sin\left(\frac{\theta_{j}}{2}\right) \ket{\bm{1}_{j}}.
\end{equation}
The set of $n_{g}$ Pauli word generators ${\hat{P}_{k}}$ are selected based on the magnitude QCC energy derivative with respect to $\tau_{k}$,
\begin{equation}\label{eq.qcc6}
  \frac{\partial E_{\text{QCC}}}{\partial \tau_{k}}\bigg\rvert_{\bm{\tau} = \bm{0}} = \frac{i}{2} \bra{\bm{\Omega}\left(\bm{\Gamma}_{\text{opt}}\right)} [\hat{P}_{k}, \hat{H}] \ket{\bm{\Omega} \left(\bm{\Gamma}_{\text{opt}}\right)},
\end{equation}
and evaluated with all $\tau_{k}$ set to zero and using an optimal set of mean-field parameters $\bm{\Gamma}_{\text{opt}}$.
Thus the QCC method requires optimization of $n_{g}$ + $2 n_{q}$ variational parameters in order to obtain the correlated wave function and total electronic energy.

Provided below is an example of how to instantiate a \mintinline{python}{VQESolver} object from a dictionary of options for the QCC ansatz, then initialize the internal objects it needs by calling \mintinline{python}{build}, and finally calling \mintinline{python}{get_resources} and \mintinline{python}{simulate} to obtain quantum resource estimations and execute the entire algorithm according to the user's options.
For this particular example, we focus on a system comprising a ring of ten equidistant hydrogen atoms $(\text{H}_{10})$, and treat the electron correlation at the QCC/minimal basis set level of theory.
Note that in the following example, we are using the \mintinline{python}{SecondQuantizedMolecule} instance \mintinline{python}{mol} that was defined in Section~{3.1}.

\begin{minted}{python}
from tangelo.algorithms import BuiltInAnsatze, VQESolver

ansatz_options = {"qcc_tau_guess": 1e-2, "deqcc_dtau_thresh": 1e-3, "max_qcc_gens": None}

vqe_solver_options = {"molecule": mol, "ansatz": BuiltInAnsatze.QCC, "qubit_mapping": "jw",
                      "initial_var_params": "random", "ansatz_options": ansatz_options}

solver = VQESolver(vqe_solver_options) 
solver.build()

vqe_resources = solver.get_resources()
optimal_vqe_energy = solver.simulate()
print(vqe_resources)
print(optimal_vqe_energy)
\end{minted}

In the previous code snippet, we inform the VQESolver to utilize the QCC ansatz from \mintinline{python}{BuiltInAnsatze} with the Jordan-Wigner encoding, and randomized variational parameters.
Additionally, the mean-field variational parameter set $\bm{\Gamma}$ is automatically selected so that $\ket{\bm{\Omega}\left(\bm{\Gamma}\right)}$ corresponds to the ground state Hartree-Fock reference wave function.
The dictionary \mintinline{python}{ansatz_options} is defined and contains options specific to the QCC ansatz class: \mintinline{python}{"qcc_tau_guess"}, \mintinline{python}{"deqcc_dtau_thresh"}, and \mintinline{python}{"max_qcc_gens"}.
The first option \mintinline{python}{"qcc_tau_guess"}, together with the value \mintinline{python}{"random"} for the option \mintinline{python}{"initial_var_params"}, results in uniform randomization of the QCC parameter set $\bm{\tau}$ over the range $[-0.01, 0.01]$.
The second option \mintinline{python}{"deqcc_dtau_thresh"} determines the threshold of the magnitude of the QCC energy gradient given by Eq.~\eqref{eq.qcc6} for determining which sets of candidate Pauli word generators should be considered for the ansatz given by Eq.~\eqref{eq.qcc4}.
The third option \mintinline{python}{"max_qcc_gens"} sets the maximum number of Pauli word generators $\hat{P}_{k}$ to utilize for constructing the QCC unitary ansatz in Eq.~\eqref{eq.qcc4}; this is the number of variational parameters in $\bm{\tau}$, $n_{g}$.
Since this option is set to \mintinline{python}{None}, one Pauli word generator from each candidate set that were characterized by QCC energy gradient magnitudes (Eq.~\eqref{eq.qcc6}) larger than \mintinline{python}{"deqcc_dtau_thresh"} will be employed to build the QCC unitary operator (Eq.~\eqref{eq.qcc4}).

The output of the calls to \mintinline{python}{get_resources} and \mintinline{python}{simulate} from the previous code snippet are shown below:
\begin{minted}[bgcolor=backgreen]{python}
{'qubit_hamiltonian_terms': 3591, 'circuit_width': 20,
 'circuit_gates': 2815, 'circuit_2qubit_gates': 1110,
 'circuit_var_gates': 185, 'vqe_variational_parameters': 185}
-5.392733992321686
\end{minted}
While a number of existing devices based on different technologies can accommodate this number of qubits (here denoted by \mintinline{python}{circuit_width}), the number of quantum gates required is far beyond what current noisy systems can faithfully execute. The number of variational parameters makes classical optimization challenging as well.\\

Overall, the features exposed by the API are designed to facilitate the exploration of different approaches, quantifying metrics of interest and partially or fully simulating an algorithm to obtain results, if relevant. The highly modular design of algorithms, combined with an API that allows fine control and easy computational resource estimation can be a powerful tool to facilitate exploration, and discovering more promising approaches. Meanwhile, classical solvers can be used to acquire reference values and get insights about the molecular system, which we can use to inform our quantum approaches. 

\subsection{Problem Decomposition}

Problem decomposition techniques can help with reducing computational resource requirements, and make use cases more tractable on current devices. The API for problem decomposition techniques is similar to what is used for quantum algorithms. Users first pass a dictionary of options used at instantiation, the \mintinline{python}{build} method then initializes all underlying objects, \mintinline{python}{simulate} runs the algorithm, and \mintinline{python}{get_resources} keeps track of computational requirements. Beyond that, each problem decomposition class offers its own additional methods. We illustrate the DMET and ONIOM methods in the examples below.

\subsubsection{DMET Example: A Ring of Ten Hydrogen Atoms}

Our H$_{10}$ use case is a strongly correlated system, which is something DMET may be appropriate for. In the example below, we aggressively decompose the system into 10 fragments of size 1 atom each. Because each fragment plays a symmetrical role, we decide to only consider one of them with a quantum approach and the others with the \mintinline{python}{CCSD} classical solver, in the snippet below. This is done purely to speed up computation and avoid duplicating calculations leading to the same results in this situation: DMET does not rely on symmetry.

\begin{minted}{python}
from tangelo.algorithms import BuiltInAnsatze
from tangelo.problem_decomposition.electron_localization import meta_lowdin_localization
from tangelo.problem_decomposition import DMETProblemDecomposition

ansatz_options = {"qcc_tau_guess": 1e-2, "deqcc_dtau_thresh": 1e-3, "max_qcc_gens": None}

dmet_options = {"molecule": mol, "verbose": False,
                "fragment_atoms": [1]*10, 
                "fragment_solvers": ["vqe"] + ["ccsd"]*9,
                "electron_localization": meta_lowdin_localization,
                "solvers_options": [{"ansatz": BuiltInAnsatze.QCC, "qubit_mapping": "scBK",
                                     "initial_var_params": "random", "up_then_down": True,
                                     "verbose": False, "ansatz_options": ansatz_options}]
                                     + [{}]*9}

dmet_solver = DMETProblemDecomposition(dmet_options)
dmet_solver.build()
\end{minted}

A closer look at the main DMET options below:
\begin{itemize}
\item
\mintinline{python}{fragment_atoms} dictates how atoms are assigned to fragments comprising our molecular system. This variable can be a list of integers (each representing the number of atoms comprising a fragment), or as a nested list of integers, where each sub-list groups together atom indices belonging in the same fragment. 

\item
\mintinline{python}{fragment_solvers} pairs each of the fragment with a solver (``ccsd'', ``vqe'', ``fci'', ...), or just applies the same solver to all fragments if only one of them is specified.

\item
\mintinline{python}{electron_localization} Available options for the electron localization scheme used, such as Meta-L{\"o}wdin localization~\cite{metalowdin} or Intrinsic Atomic Orbitals (IAO) localization~\cite{iao}.

\item
\mintinline{python}{solvers_options} provides the dictionary of options for the corresponding solver.
Here we employ the QCC method as in the previous H$_{10}$ example in conjunction with the symmetry-conserving Bravyi-Kitaev encoding.
The spin-orbitals are arranged such that all $\alpha$ spin-orbitals are indexed first followed by all $\beta$ spin-orbitals. The QCC variational parameters are initialized randomly.
\end{itemize}

The \mintinline{python}{get_resources} method shows that for a fragment containing a single hydrogen atom and treated with VQE using the options above, the resource requirements are significantly lower than a direct VQE approach for the whole system shown in Section~{3.2.2}:

\begin{minted}[bgcolor=backgreen]{python}
{'qubit_hamiltonian_terms': 9, 'circuit_width': 2,
 'circuit_gates': 11, 'circuit_2qubit_gates': 2,
 'circuit_var_gates': 1, 'vqe_variational_parameters': 1}
\end{minted}

The \mintinline{python}{simulate} method simulates the quantum algorithms on the target simulator backend (which can be specified as an option), runs the classical solvers as well, and returns the energy for this system. This provides an idea of the attainable accuracy on ``perfect'' devices, and can help us decide if a particular experiment on an actual device is relevant. This step is rather straightforward, as it assumes all relevant options have already been passed to the \mintinline{python}{dmet_solver} object:

\begin{minted}{python}
dmet_energy = dmet_solver.simulate()
print(dmet_energy)
\end{minted}
which returns 

\begin{minted}[bgcolor=backgreen]{python}
-5.401626680694919
\end{minted}

\subsubsection{ONIOM Example: Acetic Acid Interactions with Water}

\begin{figure}
    \centering
    \includegraphics{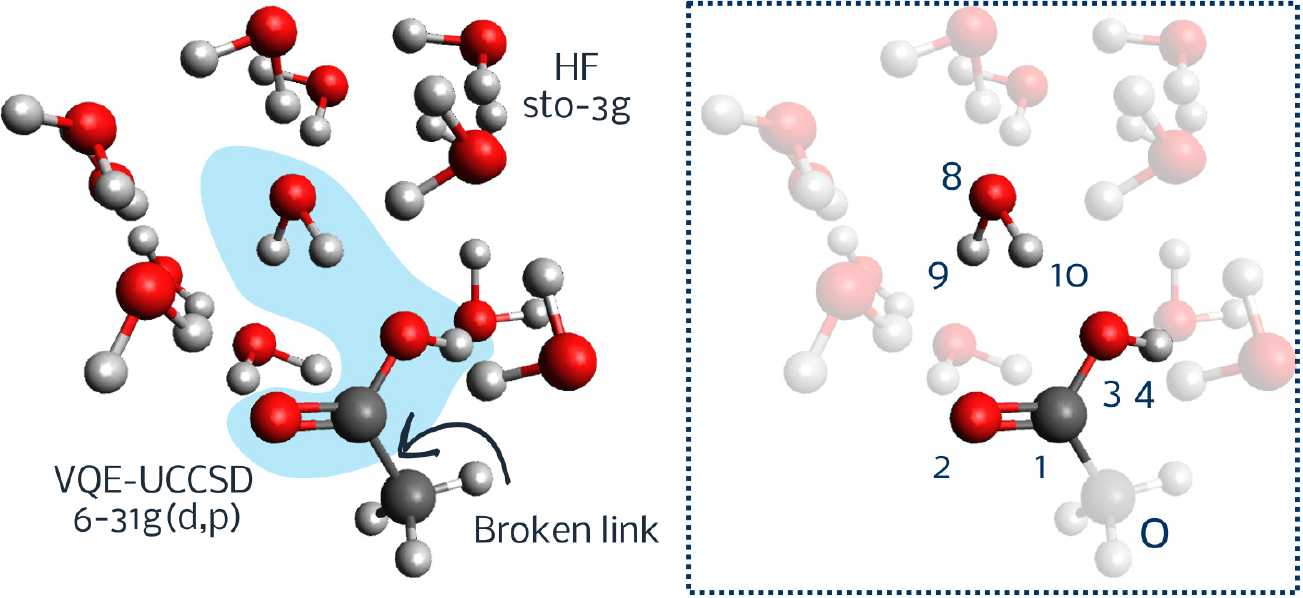}
    \caption{The acetic acid interactions with water example. The high accuracy fragment is defined using the seven numbered atoms (blue region), excluding atom 0, while the low-accuracy fragment includes the non-numbered atoms and atom 0. The broken link is repaired by adding a hydrogen atom at 0 to the high-accuracy fragment for the high accuracy calculation. Similarly a hydrogen atom at 1 is added to the low-accuracy fragment for the low-accuracy calculation.}
    \label{fig:ONIOM}
\end{figure}

Another decomposition strategy, ONIOM, may be used in cases where parts of a system can be decoupled almost completely. Here, our use case is a single acetic acid molecule positioned in a pool of water molecules: our goal is to determine the optimal heteroatom-hydrogen distance happening during solvation. The system is defined in Figure~\ref{fig:ONIOM}. Even though the system size is very far from the thermodynamic limit, computational resources remain a constraint. With the help of ONIOM, we can target a specific interaction to compute with a high-accuracy method, while considering an environment with a lower-cost electronic structure solver. This process facilitates an easier link between simulation and experimental results by keeping compute time and quantum resources practical. 

The definition of the problem is as simple as the following lines of code where \mintinline{python}{"water_system.xyz"} contains the coordinates of the system depicted in Figure~\ref{fig:ONIOM}.
\begin{minted}{python}
from tangelo.problem_decomposition.oniom import ONIOMProblemDecomposition, Fragment, Link
from tangelo.algorithms import BuiltInAnsatze as Ansatze

# Coordinates file
with open("water_system.xyz",'r') as f:
    xyz = f.read()
    # Removing first 2 lines (number of atoms and a comment line)
    xyz = xyz.split("\n", 2)[2]

options_low = {"basis": "sto-3g"}
options_high = {"basis": "6-31G**", 
                "qubit_mapping": "jw", 
                "ansatz": Ansatze.UCC3, 
                "up_then_down": True,
                "frozen_orbitals": [i for i in range(76) if i not in (16, 17)]}
                
# Whole system to be computed with a low-accuracy method (RHF, sto-3g)
system = Fragment(solver_low="rhf", options_low=options_low, charge=0)

# Fragment to be computed with a high-accuracy method (VQE-UCC3, 6-31G**).
links = [Link(0, 1, 0.709, 'H')]
model = Fragment(solver_low="rhf", options_low=options_low,
                 solver_high="vqe", options_high=options_high,
                 selected_atoms=[1, 2, 3, 4, 8, 9, 10],
                 broken_links=links,
                 charge=0)

# Construction of the ONIOM solver.
oniom_solver = ONIOMProblemDecomposition({"geometry": xyz,
                                          "fragments": [system, model],
                                          "verbose": True})
\end{minted}

The quantum resources information can be obtained with the method \mintinline{python}{get_resources}. For our chosen fragments, frozen orbitals and ansatz, the result is the following:
\begin{minted}[bgcolor=backgreen]{python}
{'qubit_hamiltonian_terms': 27, 'circuit_width': 4, 'circuit_gates': 23,
 'circuit_2qubit_gates': 8, 'circuit_var_gates': 3, 'vqe_variational_parameters': 3}
\end{minted}
The ONIOM energy can be obtained with the command \mintinline{python}{oniom_solver.simulate()}, which results in a value that is approximately $4$ Hartree lower than the Hartree-Fock solution. In a provided \href{\notebooksURL/oniom.ipynb}{example notebook}, the equilibrium O-H distance was computed, and is in agreement with reference values.

\subsection{Quantum Circuits}

Tangelo's API for defining quantum gates and circuits is both straightforward and pythonic. It is handled by the \mintinline{python}{linq} submodule, which also provides format conversion function to enable Tangelo to connect to various compute platforms --- hence the name ``linq''. The \mintinline{python}{Gate} and \mintinline{python}{Circuit} classes in Tangelo allow us to express algorithms in a backend-agnostic fashion, which can then be converted into the different formats supported by computational platforms (simulators, QPUs).\\

A \mintinline{python}{Gate} object is most accurately described as a dictionary, with a few extra methods provided for convenience. It is entirely defined by its name, indices of target and control qubits (indexing starts at 0 in Tangelo), values of parameters if any, and a ``variational'' tag in order to easily find it within a circuit, if relevant later.

\begin{minted}{python}
from tangelo.linq import Gate

# Create a Hadamard gate acting on qubit 2
H_gate = Gate("H", 2)
# Create a CNOT gate with control qubit 0 and target qubit 1
CNOT_gate = Gate("CNOT", target=1, control=0)
# Create a parameterized rotation on qubit 1 with angle 2 radians
RX_gate = Gate("RX", 1, parameter=2.)
# Create a parameterized rotation on qubit 1, with undefined angle, tag as variational
RZ_gate = Gate("RZ", 1, parameter="an expression", is_variational=True)

print(RZ_gate)
\end{minted}
outputs:
\begin{minted}[bgcolor=backgreen]{python}
RZ        target : 1   parameter : an expression    (variational)
\end{minted}

Likewise, a \mintinline{python}{Circuit} object can be pictured as a self-aware list of \mintinline{python}{Gate} objects, with a few useful methods and operator overloading. There are several ways to create and modify circuits:

\begin{minted}{python}
from tangelo.linq import Circuit

# Here's a list of abstract gates
mygates = [Gate("H", 2), Gate("CNOT", 1, control=0), Gate("CNOT", target=2, control=1),
           Gate("Y", 0), Gate("RX", 1, parameter=2.)]

# Users can create empty circuit objects and use add_gate later on
circuit1 = Circuit()
for gate in mygates:
    circuit1.add_gate(gate)
    
# Users can also directly instantiate a circuit with a list of gates (preferred)
circuit2 = Circuit(mygates)

# It is possible to concatenate circuit objects to form deeper circuits
circuit3 = Circuit(mygates) + Circuit([Gate("RZ", 4, parameter="alpha", is_variational=True)])
\end{minted}

This code snippet highlights a few interesting methods and attributes:

\begin{minted}{python}
# Examine properties of a circuit directly
print(f"The number of gates contained in circuit3 is {circuit3.size}")
print(f"The number of qubits in circuit3 is {circuit3.width}")
print(f"Does circuit have gates tagged as variational ? {circuit3.is_variational}")
print(f"Gate counts: {circuit3.counts}")

# Access and modify the variational gates (here the first one) in your circuit
circuit3._variational_gates[0].parameter = 777.
print(circuit3})
\end{minted}

outputs:
\begin{minted}[bgcolor=backgreen]{python}
"The number of gates contained in circuit3 is" 6
"The number of qubits in circuit3 is" 5
"Does circuit have gates tagged as variational ?" True
Gate counts: {'H': 1, 'CNOT': 2, 'Y': 1, 'RX': 1, 'RZ': 1}

Circuit object. Size 6 

H         target : 2   
CNOT      target : 1   control : 0   
CNOT      target : 2   control : 1   
Y         target : 0   
RX        target : 1   parameter : 2.0
RZ        target : 4   parameter : 777.0	 (variational)
\end{minted}

Some convenience operators have been defined on the \mintinline{python}{Circuit} class, such as ==, !=, + (concatenation) and * (repetition). Some additional methods are also available: 
\begin{itemize}
\item
\mintinline{python}{inverse} returns the inverse (``dagger'') of the unitary corresponding to the circuit.
\item
\mintinline{python}{split} identifies non-entangled subsystems that can be simulated independently and breaks the initial circuit into several ``narrower'' circuits, thus lowering resource requirements.
\item
\mintinline{python}{stack} combines several circuits to form a wider one, in order to fill a quantum device as much as possible. This can be used to run multiple shots in parallel, thus reducing cost of a quantum experiment. Executing it accurately can however be a more challenging task for the device.
\end{itemize}

Tangelo performs a number of checks regarding qubit indices, but ultimately allows you to provide whatever gate name or value for parameters you'd like, as long as it's one of the supported types. This data only needs to fully make sense once your circuit is converted into a backend-specific format: only then some error may arise if you are attempting to use a gate name or parameter value that is not supported by Tangelo or the target backend. The variable below returns a dictionary showing the available backends, as well as the gate names supported with each of them.
\begin{minted}{python}
from tangelo.linq import SUPPORTED_GATES
\end{minted}

\subsection{Measurement Protocols}

The number of measurements of the qubit register correlates with both the cost of an experiment and the theoretical accuracy of its results. DMET and VQE are both iterative procedures that would require a prohibitive amount of measurements. In the following, we simply focus on the steps used to recompute the total energy of the $H_{10}$ system with the DMET approach used in 3.3.1, using the circuits with optimal variational parameters and qubit operators. Extracting them from the DMET object is rather straightforward:
\begin{minted}{python}
fragment, fragment_qb_ham, fragment_circuit = dmet_solver.quantum_fragments_data[0]
\end{minted}

To obtain the 1- and 2-RDMs needed for the DMET energy calculation, we need to measure every term in the Hamiltonian that yields a qubit operator with a non-imaginary coefficient (the energy of a molecular system is real, imaginary ones would not contribute to calculations). This is where various toolboxes come to the rescue, and allow us to quickly put together custom code to identify these qubit operators. The \mintinline{python}{FermionOperator} and \mintinline{python}{QubitOperator} classes are subclasses of the classes with the same name provided by the OpenFermion package, while \mintinline{python}{fermion_to_qubit_mapping} is a unified interface to several different qubit mappings.

\begin{minted}{python}
from tangelo.toolboxes.operators import FermionOperator, QubitOperator
from tangelo.toolboxes.qubit_mappings.mapping_transform import fermion_to_qubit_mapping
    
# Find all the measurement bases that are needed to compute the RDMs.
# Accumulate them in a QubitOperator object to manipulate them afterwards.
qubit_op_rdm = QubitOperator()
bases_to_measure = set()

for term in fragment.fermionic_hamiltonian.terms:
    
    fermionic_term = FermionOperator(term, 1.0)

    qubit_term = fermion_to_qubit_mapping(fermion_operator=fermionic_term, mapping="scBK",
                                          n_spinorbitals=fragment.n_active_sos,
                                          n_electrons=fragment.n_active_electrons,
                                          up_then_down=True)
    qubit_term.compress()

    for basis, coeff in qubit_term.terms.items():
        if coeff.real != 0 and basis:
            bases_to_measure.add(basis)
            qubit_op_rdm.terms[basis] = coeff
\end{minted}

Printing \mintinline{python}{qubit_op_rdm} shows that we have identified nine terms for this $H_{10}$ ring, or equivalently nine measurement bases, leading to nine slightly different variations of our optimized quantum circuits, if we want to compute the RDMs.

\subsubsection{Term Grouping}

Applying for example the idea of qubit-wise commutativity \cite{QWC} to group terms, only five measurement bases (i.e., circuits) are needed. After a closer look, we see that the data obtained by running 5 of these circuits can be used and agglomerated to generate even better data for the remaining 4 computational bases.

\begin{figure}[H]
    \centering
    \includegraphics[scale=0.5]{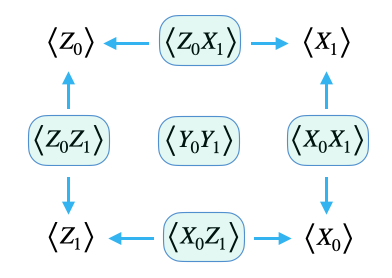}
    \caption{Grouping terms using qubit-wise commutativity tells us that collecting data for 5 measurements bases provides us with data for the 4 others, whose expectation values can be computed more accurately by combining the shots coming from several of them}
    \label{fig:qwc}
\end{figure}

Reducing the number of measurements while achieving satisfying accuracy is a challenge that increases with problem size: the number of possible terms to measure scales exponentially ($O(4^N)$ where $N$ is the number of qubits). Creating larger groups typically comes at the cost of running a longer circuit, which in most cases in unfeasible due to device noise.

The \mintinline{python}{measurements} toolbox provides tools to help us apply some of these ideas. The function below returns a dictionary where the information from measuring the parent term can be used to get the expectation value for each term in the group. This particular grouping attempts to find the Minimum Clique Cover: this problem may admit several solutions, and what is returned by the algorithm depends on the value of the $seed$ parameter, introducing some randomness.

\begin{minted}{python}
from tangelo.toolboxes.measurements import group_qwc

qwc_map = group_qwc(qubit_op_rdm, seed=0)

# Current value of qwc_map printed below
{((0, 'X'), (1, 'X')): 0.25 [X0 X1],
 ((0, 'X'), (1, 'Z')): 0.25 [X0] + 0.25 [X0 Z1] + 0.25 [Z1],
 ((0, 'Y'), (1, 'Y')): -0.25 [Y0 Y1],
 ((0, 'Z'), (1, 'X')): 0.25 [Z0] + 0.25 [Z0 X1] + 0.25 [X1],
 ((0, 'Z'), (1, 'Z')): 0.25 [Z0 Z1]}
\end{minted}

\subsubsection{Measurement Estimation}
To achieve a desired accuracy of a qubit Hamiltonian expressed as $\hat{H}=\sum_i c_i P_i$, we can create a measurement estimate for each Pauli word $P_i$ by assuming its standard deviation scales as $\frac{1}{\sqrt{n_\mathrm{shots}}}$. This is then weighted by the magnitude of its coefficient $c_i$.

\begin{minted}{python}
from tangelo.toolboxes.measurements.estimate_measurements import get_measurement_estimate

measurements = {k: get_measurement_estimate(v, digits=2) for k,v in qwc_map.items()}

# Value of qwc_map
{((0, 'X'), (1, 'X')): {((0, 'X'), (1, 'X')): 62500},
 ((0, 'X'), (1, 'Z')): {((0, 'X'),): 62500, ((0, 'X'), (1, 'Z')): 62500, ((1, 'Z'),): 62500},
 ((0, 'Y'), (1, 'Y')): {((0, 'Y'), (1, 'Y')): 62500},
 ((0, 'Z'), (1, 'X')): {((0, 'Z'),): 62500, ((0, 'Z'), (1, 'X')): 62500, ((1, 'X'),): 62500},
 ((0, 'Z'), (1, 'Z')): {((0, 'Z'), (1, 'Z')): 62500}}
\end{minted}

Beyond the examples above, emergent methods like classical shadows~\cite{ClassicalShadow} are developed to mitigate the measurement overhead by offloading quantum tasks to the pre- and post-processing steps. This prediction protocol exhibits logarithmic scaling with the number of shots to evaluate properties of a state. Tangelo users can leverage this protocol by performing the relevant pre- and post-processing functions. We look forward to introducing more grouping techniques and data-structures to help users devise efficient measurement protocols within their target budget.

\subsection{Executing Quantum Circuits}

Tangelo offers several ways to submit an experiment to a quantum device or a noisy simulator that can approach realistic behavior. The output of this step leads to a probability distribution, mapping bitstrings to how often they have been observed for the specified number of shots. Depending on the platform used to run the experiment, some additional data about the hardware and how it performed may be available.

\subsubsection{Using QEMIST Cloud}

The simplest way to submit an experiment to any device available on the supported quantum cloud services is through \href{\QEMISTCloudURL}{QEMIST Cloud}'s client library, allowing users to run quantum hardware experiments using their QEMIST Cloud account credentials and credits. This is possible if you have access to QEMIST Cloud, and have installed the qemist-client Python package. For more details about this feature, please refer to our dedicated \href{\notebooksURL/qemist_cloud_hardware_experiments_braket.ipynb}{notebook}, and \href{\GCcontactURL}{reach out} about QEMIST Cloud. Our convenience wrappers and functions can facilitate the submission and cost estimation of experiments.

Below, a simple code snippet illustrating how to run 10,000 shots of the YY circuit on IonQ's hardware, through Amazon's Braket quantum cloud services:

\begin{minted}{python}
# Retrieve both these values from your QEMIST Cloud dashboard
import os
os.environ['QEMIST_PROJECT_ID'] = "your_project_id_string"
os.environ['QEMIST_AUTH_TOKEN'] = "your_qemist_authentication_token"

from tangelo.linq.qpu_connection import job_submit, job_status, job_result, job_estimate

# Estimate cost of task on different backends
price_estimates = job_estimate(circuit_YY, n_shots=n_shots)

# Submit task and retrieve results
backend = 'arn:aws:braket:::device/qpu/ionq/ionQdevice'
job_id = job_submit(circuit_YY, n_shots=n_shots, backend=backend)
freqs, raw_data = job_result(job_id)
\end{minted}

\subsubsection{Using a Quantum Cloud API and Format Conversion}

The utility functions in \mintinline{python}{tangelo.linq} allow us to convert our generic circuit object into a variety of formats supported by other open-source packages and services, such as Amazon's Braket and Microsoft's Azure Quantum. You can thus convert a Circuit object into the desired format and use the API of those services directly in order to reach a QPU or an online simulator.

If you convert a Circuit object into the Braket format, the submission process is pretty straightforward provided that you have a Braket account, as demonstrated in the Braket documentation. Likewise, Azure Quantum supports a number of formats. Our package provides similar ``translation'' functions allowing us to produce a Circuit object into the Q\#, qiskit or a cirq format, for example. Provided we have a working Azure Quantum environment, our circuits can be submitted using their API. The code snippet below illustrate how straightforward the process is:

\begin{minted}{python}
from tangelo.linq.translator import translate_braket, translate_cirq, translte_qiskit
braket_circuit = translate_braket(circuit_YY)
cirq_circuit = translate_cirq(circuit_YY)
qiskit_circuit = translate_qiskit(circuit_YY)
\end{minted}

Some hardware providers rely on a specific format and may provide an API for directly submitting experiments to their devices to their close collaborators. The format conversion functions can be used in all these cases as well; check out what your favorite hardware providers offer, we are happy to support new platforms users deem relevant.

\subsubsection{Emulation on a Simulator Backend}

The format conversion functions can be used to target the simulators of your choice, and are also useful if you need to save or share a circuit in a particular format with your collaborators. The \mintinline{python}{linq} module provides the \mintinline{python}{Simulator} class, which acts as a unified interface to the various backends we support. You can modify the behaviour of your \mintinline{python}{Simulator} object by indicating the target backend, number of shots, or noise model.

The convenient \mintinline{python}{simulate} method allows you to immediately simulate your circuit: you do not have to write any backend-specific code or explicitly use the format conversion functions. This is done automatically for you, you only have to specify a backend.

\begin{minted}{python}
# Noiseless simulator returning the state vector
sim_cirq = Simulator(target="cirq")
freqs, sv = sim_cirq.simulate(my_circuit, return_statevector=True))

# Noiseless simulator running multiple shots
sim_qulacs = Simulator(target="qulacs", n_shots=1000)
freqs, _ = sim_qulacs.simulate(my_circuit}
\end{minted}

The simulate method returns a 2-tuple:
\begin{itemize}
\item 
A sparse histogram of frequencies associated to the different observed states, in least-significant qubit first order (e.g '01' means qubit 0 (resp. 1) measured in $|0\rangle$ (resp. $|1\rangle$) state). That is, it is to be read ``left-to-right'' in order to map each qubit to the basis state it was observed in.

\item 
A representation of the quantum state (often a state vector), if available on the target backend and if the user specifies it using the \mintinline{python}{return_statevector} optional parameter. Otherwise, this returns \mintinline{python}{None}.
\end{itemize}

While we return frequency histogram keys in a standardized way, we do not alter the internal representation of the quantum state provided by the target backend. We however provide some information in variable \mintinline{python}{linq.backend_info}, like the ordering of the entries of the state vector for example, which is helpful if you use the \mintinline{python}{initial_statevector} option of \mintinline{python}{simulate}, to avoid re-simulating a sequence of gates leading to a known state vector.

Below is an example of how you can use the generic \mintinline{python}{NoiseModel} object to perform noisy simulation. We show an example applying a depolarization channel to specific gates, each with a given probability.

\begin{minted}{python}
from tangelo.linq import Simulator
from tangelo.linq.noisy_simulation import NoiseModel

nmp = NoiseModel()
nmp.add_quantum_error("CNOT", "depol", 0.01)
nmp.add_quantum_error("RZ", "depol", 0.005)
nmp.add_quantum_error("H", "depol", 0.005)
backend_noisy = Simulator(target="cirq", n_shots=n_shots, noise_model=nmp)
\end{minted}

\subsection{Post-processing}

\subsubsection{Expectation Values}

The \mintinline{python}{Simulator} class provides two methods to compute expectation values with regards to an object of type \mintinline{python}{QubitOperator}.

The \mintinline{python}{Simulator} class method  \mintinline{python}{get_expectation_values_from_frequencies_oneterm} computes expectation values from a histogram of frequencies and a single-term qubit operator. This is useful for post-processing the outcome of a quantum circuit executed on a device, or when separating simulation from post-processing. By looping over the terms and corresponding frequencies, users are able to compute the expectation values of qubit operators with an arbitrary number of terms.

If you are using a simulator, the \mintinline{python}{get_expectation_value} method is the fastest way to simulate a circuit and compute its expectation value with regards to a qubit operator using a \mintinline{python}{Circuit} object.

\begin{minted}{python}
# Openfermion operators can be used
from openfermion.ops import QubitOperator
op = 1.0 * QubitOperator('Z0')

# Directly through a simulator backend, providing the state-preparation circuit
# A single line regardless of the size of the qubit operator
sim = Simulator(target="cirq")
expval1 = sim.get_expectation_value(op, c)

# Or from a histogram computed separately beforehand, with a single-term qubit operator.
# Useful for post-processing results of an experiment on a quantum device
freqs, _ = sim.simulate(c) # This could happen on a quantum device instead
term, coef = tuple(op.terms.items())[0]  # This yields ((0, 'Z'),), 1.0
expval2 = coef * Simulator.get_expectation_value_from_frequencies_oneterm(term, freqs)
\end{minted}

\subsubsection{Error Mitigation}

Due to noise, the hardware produces a mixed state, which reduces the accuracy of our observables. Although error correction is not currently available, we can use hardware-agnostic post-processing techniques to mitigate noise on near-term hardware. Tangelo aims to provide, through its toolboxes, a collection of noise-mitigation techniques.

As an example of error mitigation for the DMET $H_{10}$ ring use case, we use a density matrix purification technique based on McWeeny's method~\cite{2rdmpurif} to purify our noisy state to the dominant eigenvector. For our particular use case, we can use this method for the 2-RDM since our fragments consist of two electrons --- thus the 2-RDM is the full density matrix, and idempotency can be imposed.  We can draw the required function from a toolbox and apply it as follows:

\begin{minted}{python}
from tangelo.toolboxes.post_processing.mc_weeny_rdm_purification import mcweeny_purify_2rdm
onerdm, twordm = mcweeny_purify_2rdm(twordm, conv=1e-2)
\end{minted}

This method is however limited to systems with two electrons: in general, applying the technique to 2-RDMs of higher electron systems would require the more sophisticated N-representability conditions. Application-agnostic error mitigation methods such as error extrapolation are also available in Tangelo~\cite{DIIS,Richardson,Richardson_analytical}. We are working on integrating more error-mitigation techniques, chemistry-inspired or not.

\subsection{Statistical Treatment}

Experimental data requires a measure of its uncertainty. As it is often prohibitively expensive to collect large amounts of data on quantum computers for the purpose of estimating uncertainty, we generate statistics from our dataset using an established method called bootstrapping~\cite{bootstrap}. For each histogram obtained from our experiment, we resample with replacement from that distribution to generate new histograms of the same sample size. We then use these histograms to calculate a new set of expectation values, RDMs, and total energies. This process is repeated many times to form a statistical series, from which we calculate the average energy and standard deviation of our experiment.

We can put this together rather quickly by drawing the \mintinline{python}{get_resampled_frequencies} function in one of the toolboxes in Tangelo, and combining it with others to arrive at the result.

\begin{minted}{python}
from tangelo.toolboxes.post_processing.bootstrapping import get_resampled_frequencies

fragment_energies = list()

for n in range(10000): # 10000 samples
    
    # Draw random bootstrap samples and construct new histograms.
    resample_freq = {term: get_resampled_frequencies(freq, n_shots) 
                     for term, freq in freq_dict.items()}
    
    # Compute expectation values.
    exp_vals = {term: Simulator.get_expectation_value_from_frequencies_oneterm(term, hist)
                for term, hist in resample_freq.items()}
    
    # Construct 1- and 2-RDMs, purify with Mcweeny.
    onerdm, twordm, _, _ = compute_rdms(fragment, exp_vals)
    onerdm_spinsum, twordm_spinsum = mcweeny_purify_2rdm(twordm, conv=1e-2)
    
    # Calculate the total energy.
    e_fragment = compute_electronic_fragment_energy(fragment, onerdm_spinsum, twordm_spinsum)
               + core_constant
    fragment_energies.append(e_fragment)
    
# Calculate the mean and standard deviation.
mean = np.mean(fragment_energies)
stdev = np.std(fragment_energies, ddof=1)
\end{minted}

In this situation, the reusable code found in the various toolboxes in Tangelo can help us quickly draw what we need, and put together code to arrive at the relevant answers.

\section{Open-source Management, Community and Philosophy}

Tangelo was designed as a tool to empower both its developers and the community at large. The field of quantum computing is seeing a lot of innovation and moving at a fast pace: we decided to make this software open-source and compatible with multiple frameworks to enable the community to make the most of this momentum.

By developing Tangelo, we aspire to support users from both academia and industry in designing successful quantum chemistry applications and experiments on quantum devices. We hope to build a community around these outcomes, and further develop this package with the contributions of our users, to the benefit of the field as a whole.

\subsection{Software Development Practices}

Tangelo is a package written in Python 3, following the PEP8 coding guidelines~\cite{pep8}. It provides extensive documentation and descriptive names for classes and functions, in order to support usage and further developments from the community. We rely on the Python unittest framework in order to systematically test that the code works as intended, and guarantee reliable outcomes.

We strive to develop code that is easy to use, and as modular as possible in order to provide the community with highly-reusable building blocks that can be assembled or tinkered with, to support our quantum explorations.

Tangelo is hosted on Github, at \href{\githubURL}{\githubURL}. Users can contribute in various ways detailed in the \href{\githubURL/blob/main/CONTRIBUTIONS.rst}{contribution file}, including making feature requests and reporting bugs through the Issue tab. It is possible for anyone to contribute code to this project by following the standard Pull Request (PR) process available on Github, which is reviewed by at least one of the members of the Tangelo team before being integrated into the code base. In order to ensure that Tangelo remains reliable, the package is automatically rebuilt, tested, and the documentation updated after a pull request has been merged.

\subsection{Distribution}

There are several different channels of distribution for Tangelo. Tangelo can be deployed over MacOS and Linux in a straightforward way using tools such as git and pip, and also to Windows, using the Windows Linux Subsystem (wsl), or Docker.

User can directly retrieve the source code of Tangelo on Github. This option is better suited for users who want to benefit from the latest changes and ongoing developments on the various branches of the repository, as well as work collaboratively. It is then possible to install the package and its dependencies using pip, from the top directory of Tangelo, containing the file \verb|setup.py|. 

Please refer to the latest installation instructions in the Github repository (\href{\githubURL/blob/main/README.rst}{README file}), as the following instructions may be outdated in the future.

Tangelo is distributed through the Python Package Index (PyPI)~\cite{pypi}, which means users can directly install it through the command
\begin{minted}{bash}
  $ pip install tangelo-gc
\end{minted}

After cloning the repository or downloading the files from Github, it is possible to install Tangelo from sources. This may be helpful to developers, as well as users who wish to deploy Tangelo in specific environments, such as High-Performance Computing systems or systems with no access to PyPI.
\begin{minted}{bash}
  $ python -m pip install .
\end{minted}

Finally, Tangelo can be deployed to various environments using Docker. The GitHub repository contains a file called \verb|Dockerfile| which can be used to build a Docker container and deployed in any environment, including Windows.\\

Tangelo features a number of optional dependencies that can be installed separately (qiskit, qulacs, braket...), following the instructions of their respective development teams. Most of them come in the form of Python packages available on PyPI, and are straightforward to install. A number of these packages are related to the various quantum circuit simulators and formats supported.

\section{Closing remarks}

It is our wish to expand the capability of this platform and develop a community around it, to further research applications of quantum computing to materials science as well as providing the tools to design successful hardware experiments.

The field offers an abundance of challenges to tackle, and we need to leverage the skills of people with various backgrounds to overcome them. Although the primary areas of interest seem to be quantum computing and advanced materials simulation, various domains of applied mathematics and software development play a crucial role in the pace of innovation. You do not necessarily need to be a seasoned software-developer or a quantum computing expert to contribute to this project. By sharing your ideas and your developments with the community, you are creating an opportunity for us to learn and grow together, and take ideas to the finish line and beyond.

\vspace{4mm}
Please do not hesitate to reach out to report bugs or suggest new features for Tangelo. Implementation of state-of-the-art quantum chemistry algorithms, functions helping users to connect to new backends, reducing measurement overhead, approaches based on first quantization, and computing other properties of materials are just a few of the many topics that are relevant to the community. Integration with other packages is also desirable, including those focusing on quantum circuit optimization, compilation, and simulation.\\

What will you do with Tangelo?

\section{Acknowledgments}

The authors would like to thank our advisor Isaac Kim (UC Davis) for his continuous input and feedback. We acknowledge the technical contributions from the ongoing research collaboration between the Good Chemistry Company and Dow, Inc., which helped shape the foundation of Tangelo. We would also like to thank Nima Alidoust for his advice and vision.

We would like to thank all the contributors to the various open-source projects this software and our research rely on. Tangelo leverages various python packages, focusing on topics such as classical chemistry, quantum computing applied to materials science, or quantum circuit emulators and interfaces connecting to quantum devices, among others. \cite{pyscf1,pyscf2,Cirq,Qiskit,qdk,Amazon_Braket,Qulacs,OpenFermion}
\printbibliography

\end{document}